# Shared optical wireless cells for in-cabin aircraft links

Osama Zwaid Alsulami[1], Sarah O. M. Saeed[1], Sanaa Hamid Mohamed[1],
T. E. H. El-Gorashi[1], Mohammed T. Alresheedi[2] and Jaafar M. H. Elmirghani[1]
[1]School of Electronic and Electrical Engineering, University of Leeds, LS2 9JT, United Kingdom
[2]Department of Electrical Engineering, King Saud University, Riyadh, Kingdom of Saudi Arabia
ml15ozma@leeds.ac.uk, elsoms@leeds.ac.uk, elshm@leeds.ac.uk,
t.e.h.elgorashi@leeds.ac.uk, malresheedi@ksu.edu.sa, j.m.h.elmirghani@leeds.ac.uk

**ABSTRACT**
The design of a wireless communication system that can support multiple users at high data rates inside an aircraft is a key requirement of aircraft manufacturers. This paper examines the design of an on-board visible light communication (VLC) system for transmitting data on board Boeing 747-400 aircraft. The reading light unit of each seat is utilised as an optical transmitter. A red, yellow, green, and blue (RYGB) laser diode (LD) is used in each reading light unit for transmitting data. An angle diversity receiver (ADR), which is an optical receiver that is composed of four branches (in this work), is evaluated. The signal-to-interference-plus-noise ratio (SINR) and data rate are determined. Three scenarios have been examined where, in the first scenario, one device is used, in the second scenario two devices are used and in the third scenario three devices are used by each passenger. The proposed system can offer high SINRs that support high data rates for each passenger by using simple on-off-keying (OOK).

## 1. INTRODUCTION

There is growing demand among aircraft manufacturers for transferring data at high data rates with decreased interference from devices inside the cabin of the aircraft using wireless communication. These demands must be supported by state of the art wireless communication technologies. Current wireless communication technology uses radio frequency (RF) to transmit data. The RF system has several limitations, such as channel capacity, limited achievable transmission rate and interference with other systems within the aircraft and the use of the limited available radio spectrum. Achieving very high data rates above 10 Gbps whilst supporting multiple users can be difficult when using the radio spectrum [1]–[8]. The number of internet users is expected to increase by 27 times in the five years between 2016 and 2021 [9]. Optical wireless communication (OWC) is a promising future technology for wireless communication. It offers many advantages, such as high security, license free bandwidth and cheaper components, compared to RF systems [10]–[18]. OWC systems have been investigated in many studies showing that video, data and voice can be transmitted at data rates of up to 25 Gbps by using OWC systems in indoor environments [7], [8], [10], [26], [27], [19]–[25], [28]. Thus it is a potential solutions that can be used inside the cabin of aircrafts to offer connections between passengers and the Internet. Furthermore, it can provide a solution for the incompatibility of aircraft instruments with RF. Many studies have applied a variety of techniques to reduce interference between devices [4], [29]–[33]. Wavelength division multiple access (WDMA) can help reduce interference and increase the signal-to-interference-plus-noise ratio (SINR) [4]. In addition, optimised resource allocation can also help increase the system performance by reducing interference between devices.

In this paper, a visible light communication (VLC) system, which is a type of OWC system, is used for communication on-board the aircraft. The VLC transmitter that is used in this work has red, yellow, green, and blue (RYGB) Laser Diodes (LDs) that can offer several advantages, such as white colour for the purpose of illumination in the indoor environment as shown in [31] and high modulation bandwidth for communication. An angle diversity receiver (ADR) is the optical receiver that is evaluated in this work. In addition, WDMA is considered in this work to support multiple access. Mixed Integer Linear Programme (MILP) is used to optimise the resource allocation. The rest of this paper is organised as follows: Section 2 describes the system configuration. The design of the optical receiver is illustrated in Section 3. Section 4 provides the simulation results and the conclusions are provided in Section 5.

## 2. SYSTEM CONFIGURATION

The Boeing 747-400 aircraft cabin is the environment in which this system is evaluated. The design considered only the aircraft economy class. This class consists of up to 539 passenger seats, as shown in Figure 1a. The cabin dimensions are 57 m length × 6.37 m width × 2.41 m height, based on [34]. The cabin was divided into small sections in the simulation for the purpose of simplifying the computation time. In this study, up to second order reflections were examined as higher order reflections have little impact on the received signal [35]. A ray-tracing algorithm is used in Matlab to calculate reflections from the ceiling, walls and floor of the cabin. Thus, cabin surfaces were divided into small equals areas in the first and second order reflections with a given reflection coefficient. Each surface is considered to have reflections similar to those of a plaster wall which reflects signals similar to a Lambertian pattern as determined in [36]. The reflection coefficient of the ceiling and walls is set to 0.8, while 0.3 is set for all other surfaces, including the floor. The small equal areas in each surface reflect signals in the form of a Lambertian pattern. These small areas of the surfaces have an inverse relationship with the resolution of the result. When small areas are used for the reflecting elements in the simulations, the resolution of the results increases, but the simulation computation time also increases. Thus, to keep a balance between the resolution of the results and the simulation computation time, the surface area is divided into 5 cm × 5 cm relecting elements for the first order reflection and 20 cm × 20 cm for the second order reflection [5], [17]. The reading light unit placed above each seat is used for communication and illumination. Each reading light unit consists of two RYGB LDs with a narrow beam to cover one seat in the cabin. The size of the beam of each reading light unit can be selected based on the Lambertian emission order ($n$). Thus, when $n$ is increased, the beam size decreases, reducing the coverage area of the beam. As a result, intersection will be reduced between reading light units to reduce interference. The beam semi angle is set at $14°$ for the three reading light units that are placed at the top of the three seats next to the window (seats nos. 1, 2, 3, 8, 9, 10), as shown in Figure 1 to cover one seat. While, the beam semi angle of the other reading light units that are placed at the top of the 4 middle seats (seats nos.: 4, 5, 6, 7), as shown in Figure 1 is set at 10° to lead each reading light unit to cover one seat. All communications occur above the seats' surfaces which means communications below the surfaces of the seats are blocked.

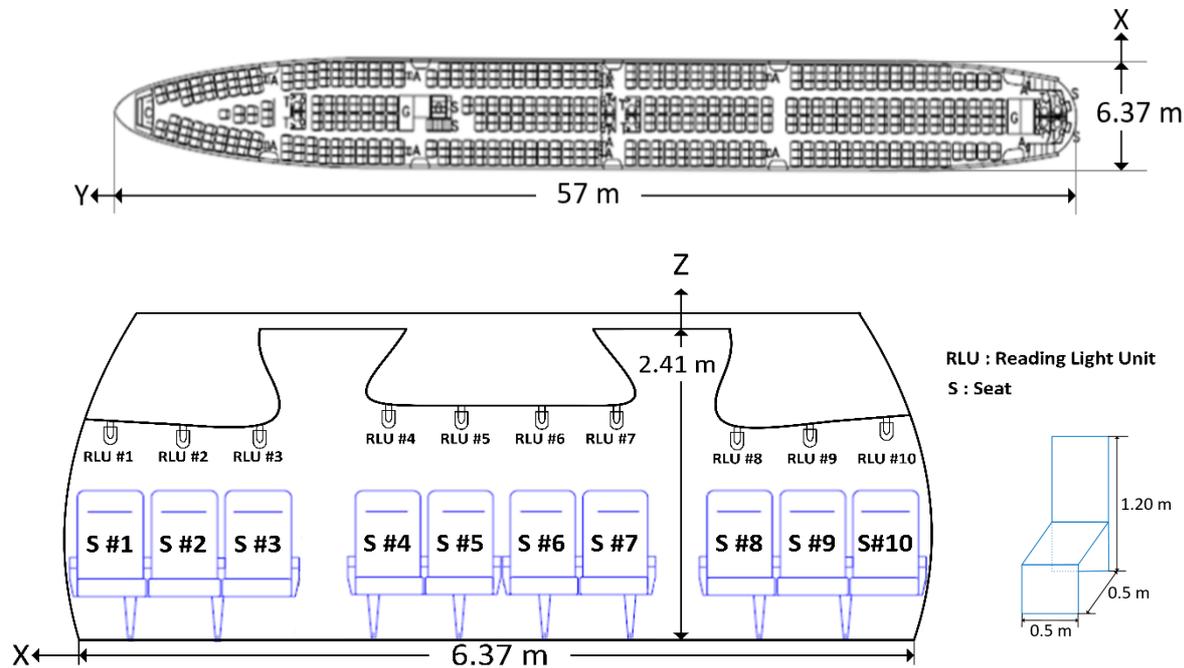

Figure 1: System Configuration

## 3. OPTICAL RECEIVER DESIGN

An angle diversity receiver (ADR), similar to [7], was evaluated in this work. The ADR consists of 4 detectors, as shown in Figure 2, to collect signals and reduce interference and each detector is oriented to cover different areas with a narrow field of view (FOV). The orientation of each detector is based on two angles; the elevation ($El$) and the azimuth ($Az$) angles. The $El$ angle for all detectors is set equal to 70° and the $Az$ angles are set at 45°, 135°, 225° and 315°. The FOV of each detector was chosen to be 21°. The area of each detector is set at $10\ mm^2$ and the responsivity of each wavelength is 0.4 A/W for red, 0.35 A/W for Yellow, 0.3 A/W for green 0.2 A/W for blue wavelengths.

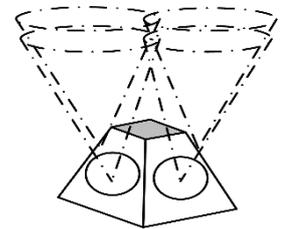

Figure 2: ADR

## 4. SIMULATION SETUP AND RESULTS

In this work, downlink communications of three scenarios were evaluated, where each passenger uses one, two or three devices, such as mobile phone, laptop or smartwatch at the same time. An MILP model similar to [28] and [37] was developed building on our track record in MILP in [38] to allocate each device a different wavelength to maximise the SINR. Note that the work can be extended to cases where the transmitter and receiver are both on the communication floor [39] and can be extended to consider the uplink case [40]. The interference from different devices of one passenger and form other passengers' devices have been considered in this work. A controller that is placed inside each reading light unit already has the device location of each passenger, similar to the work in [28]. Optimising the allocation of the light unit and wavelength can enhance the system's proformance by reducing interference and increasing the sum SINR.

Table 1 shows the wavelength and reading light unit allocation for scenario one, where each passenger uses one device. Table 2 illustrates the resource unit allocation for scenario two, where each passenger uses two devices at the same time. In Table 3, the optimised allocation of scenario three is shown, where each passenger utilises three devices simultaneously.

Table 1: The optimised resource allocation for scenario 1.

| Device # | Passenger #1 | | | Passenger #2 | | | Passenger #3 | | |
|---|---|---|---|---|---|---|---|---|---|
| | Reading Light Unit # | Branch # | Wavelength | Reading Light Unit # | Branch # | Wavelength | Reading Light Unit # | Branch # | Wavelength |
| 1 | 1 | 3 | All | 2 | 4 | All | 3 | 2 | All |

| Device # | Passenger #4 | | | Passenger #5 | | | Passenger #6 | | | Passenger #7 | | |
|---|---|---|---|---|---|---|---|---|---|---|---|---|
| | Reading Light Unit # | Branch # | Wavelength | Reading Light Unit # | Branch # | Wavelength | Reading Light Unit # | Branch # | Wavelength | Reading Light Unit # | Branch # | Wavelength |
| 1 | 4 | 3 | All | 5 | 1 | All | 6 | 4 | All | 7 | 2 | All |

| Device # | Passenger #8 | | | Passenger #9 | | | Passenger #10 | | |
|---|---|---|---|---|---|---|---|---|---|
| | Reading Light Unit # | Branch # | Wavelength | Reading Light Unit # | Branch # | Wavelength | Reading Light Unit # | Branch # | Wavelength |
| 1 | 8 | 2 | All | 9 | 4 | All | 10 | 3 | All |

Table 2: The optimised resource allocation for scenario 2.

| Device # | Passenger #1 | | | Passenger #2 | | | Passenger #3 | | |
|---|---|---|---|---|---|---|---|---|---|
| | Reading Light Unit # | Branch # | Wavelength | Reading Light Unit # | Branch # | Wavelength | Reading Light Unit # | Branch # | Wavelength |
| 1 | 1 | 2 | Red | 2 | 4 | Red | 3 | 1 | Yellow |
| 2 | 1 | 4 | Yellow | 2 | 3 | Yellow | 3 | 3 | Red |

| Device # | Passenger #4 | | | Passenger #5 | | | Passenger #6 | | | Passenger #7 | | |
|---|---|---|---|---|---|---|---|---|---|---|---|---|
| | Reading Light Unit # | Branch # | Wavelength | Reading Light Unit # | Branch # | Wavelength | Reading Light Unit # | Branch # | Wavelength | Reading Light Unit # | Branch # | Wavelength |
| 1 | 4 | 3 | Red | 5 | 2 | Yellow | 6 | 4 | Red | 7 | 1 | Yellow |
| 2 | 4 | 4 | Yellow | 5 | 1 | Red | 6 | 3 | Yellow | 7 | 2 | Red |

| Device # | Passenger #8 | | | Passenger #9 | | | Passenger #10 | | |
|---|---|---|---|---|---|---|---|---|---|
| | Reading Light Unit # | Branch # | Wavelength | Reading Light Unit # | Branch # | Wavelength | Reading Light Unit # | Branch # | Wavelength |
| 1 | 8 | 1 | Yellow | 9 | 4 | Red | 10 | 2 | Red |
| 2 | 8 | 3 | Red | 9 | 3 | Yellow | 10 | 4 | Yellow |

Table 2: The optimised resource allocation for scenario 2.

| Device # | Passenger #1 | | | Passenger #2 | | | Passenger #3 | | |
|---|---|---|---|---|---|---|---|---|---|
| | Reading Light Unit # | Branch # | Wavelength | Reading Light Unit # | Branch # | Wavelength | Reading Light Unit # | Branch # | Wavelength |
| 1 | 1 | 3 | Red | 2 | 1 | Red | 3 | 2 | Red |
| 2 | 1 | 3 | Green | 2 | 3 | Yellow | 3 | 3 | Green |
| 3 | 1 | 4 | Yellow | 2 | 4 | Green | 3 | 4 | Yellow |

| Device # | Passenger #4 | | | Passenger #5 | | | Passenger #6 | | | Passenger #7 | | |
|---|---|---|---|---|---|---|---|---|---|---|---|---|
| | Reading Light Unit # | Branch # | Wavelength | Reading Light Unit # | Branch # | Wavelength | Reading Light Unit # | Branch # | Wavelength | Reading Light Unit # | Branch # | Wavelength |
| 1 | 4 | 2 | Red | 5 | 4 | Red | 6 | 1 | Red | 7 | 3 | Red |
| 2 | 4 | 4 | Green | 5 | 4 | Yellow | 6 | 4 | Green | 7 | 4 | Yellow |
| 3 | 4 | 3 | Yellow | 5 | 3 | Green | 6 | 3 | Yellow | 7 | 3 | Green |

| Device # | Passenger #8 | | | Passenger #9 | | | Passenger #10 | | |
|---|---|---|---|---|---|---|---|---|---|
| | Reading Light Unit # | Branch # | Wavelength | Reading Light Unit # | Branch # | Wavelength | Reading Light Unit # | Branch # | Wavelength |
| 1 | 8 | 2 | Red | 9 | 1 | Red | 10 | 3 | Red |
| 2 | 8 | 3 | Green | 9 | 3 | Yellow | 10 | 3 | Green |
| 3 | 8 | 4 | Yellow | 9 | 4 | Green | 10 | 4 | Yellow |

Figures 3, 4 and 5 illustrate the SINR and data rate of the three scenarios. A high data rate with high SINR can be obtained by each passenger, even where three devices are used at the same time. In addition, at least one device per passenger can have a high data rate beyond 10 Gbps in all scenarios. In scenario 1, where each passenger uses one device a high SINR is observed which can support a high data rate between 15 Gbps and 22 Gbps, as shown in Figure 3. In scenario 2, where each passenger uses two devices, the SINR is still high but some device locations of some passengers such as device 2 of passenger 1 cannot support a data rate beyond 7 Gbps due to its channel bandwidth as shown in Figure 4. While, the key issue in scenario 3 is the interference between the passengers' devices which limits the data rate to around 5 Gbps for some devices.

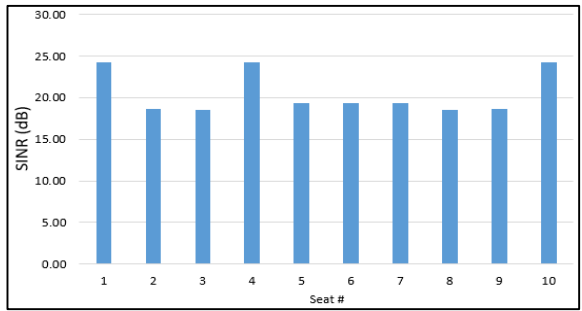 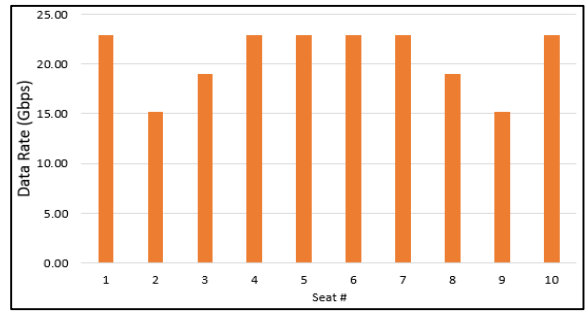

(a)                                                 (b)

Figure 3: Scenario 1 (a) SINR, (b) Data Rate.

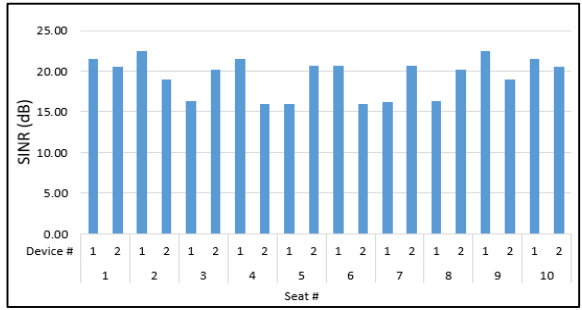 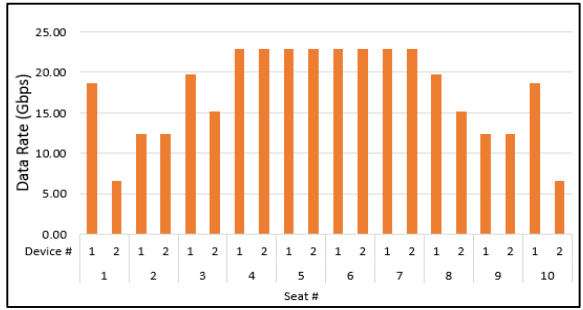

(a)                                                 (b)

Figure 4: Scenario 2 (a) SINR, (b) Data Rate.

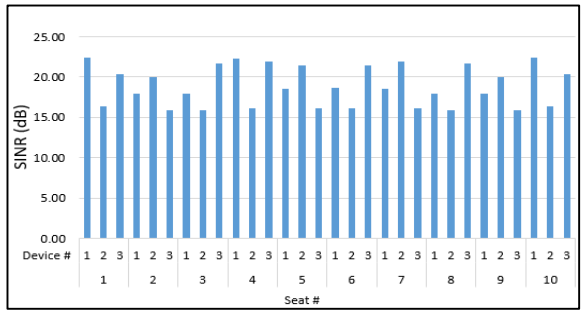 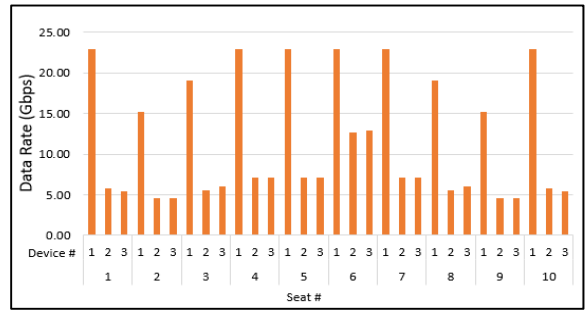

(a)                                                 (b)

Figure 5: Scenario 3 (a) SINR, (b) Data Rate.

## 5. CONCLUSIONS

An on-board VLC system for boing 747-400 was proposed in this work. The proposed system utilises the reading light units to transfer data to the passenger. An angle diversity receiver that consists of four branches was used as an optical receiver. The SINR and data rate for each device per passenger has been determined in this work. Three scenarios were examined: In scenario 1 each passenger uses one device, in scenario 2 each passenger uses two devices and in scenario 3 each passenger uses three devices. The proposed system can offer a high SINR and high data rate for different passengers at different locations inside the aircraft by using OOK modulation. A high SINR that supports a high data rate was obtained for each passenger with one, two or three devices.

**ACKNOWLEDGEMENTS**

The authors would like to acknowledge funding from the Engineering and Physical Sciences Research Council (EPSRC) INTERNET (EP/H040536/1), STAR (EP/K016873/1) and TOWS (EP/S016570/1) projects. The authors extend their appreciation to the deanship of Scientific Research under the International Scientific Partnership Program ISPP at King Saud University, Kingdom of Saudi Arabia for funding this research work through ISPP#0093. OZA would like to thank Umm Al-Qura University in the Kingdom of Saudi Arabia for funding his PhD scholarship. SOMS would like to thank the University of Leeds and the Higher Education Ministry in Sudan for funding her PhD scholarship. SHM would like to thank EPSRC for providing her Doctoral Training Award scholarship. All data are provided in full in the results section of this paper.